\documentclass[aps,prd,superscriptaddress,showpacs,twocolumn,preprintnumbers,showkeys,superscriptaddress,amsmath,amssymb,nofootinbib]{revtex4-2}
\usepackage{array}
\usepackage{booktabs}
\usepackage{commath}
\usepackage{graphicx}
\usepackage{epstopdf}
\usepackage{amsmath}
\usepackage{amsfonts}
\usepackage{amssymb}
\usepackage{latexsym}
\usepackage{hyperref}
\usepackage[english]{babel}
\usepackage[utf8]{inputenc}
\usepackage[colorinlistoftodos]{todonotes}
\usepackage{xcolor}
\usepackage{slashed}
\usepackage{bm}  
\usepackage[normalem]{ulem} 
\usepackage{amsmath,mathtools}
\usepackage{caption}
\usepackage{subcaption}
\usepackage{float}
\useunder{\uline}{\ul}{}
\usepackage{hyperref}
\usepackage{dsfont}
\usepackage{layout}
\begin{document}
%

%
\title{First-order phase-transition on dynamical Lorentz symmetry breaking system}
\author{Y. M. P. Gomes}\email{yurimullergomes@gmail.com}
\affiliation{Departamento de Física Teórica, Universidade do Estado do Rio de Janeiro, 20550-013, Rio de Janeiro, RJ, Brazil}
\author{M. J. Neves}\email{mariojr@ufrrj.br}
\affiliation{Departamento de F\'isica, Universidade Federal Rural do Rio de Janeiro,
BR 465-07, 23890-971, Serop\'edica, RJ, Brazil}

\begin{abstract}
A model of $N$ 4-component massless fermions in a quartic self-interaction based on ref. \cite{gomes2022} is investigated in the presence
of chemical potential and temperature via optimized perturbation theory that accesses finite-N contributions. We use the generating functional approach to calculate the corrections to the effective potential of the model. The model introduces an auxiliary pseudo-vector field with a nontrivial minimum and is influenced by temperature $(T)$ and chemical potential $(\mu)$. These thermodynamic quantities are introduced through Matsubara formalism. Thereby, the integrals are modified, and via the principle of minimum sensitivity, we obtain the gap equations of the model. The correspondent finite-N solutions of these equations define the vacuum states of the model associated with the background pseudo-vector field. In particular, one focuses on its temporal component that acts as an effective chiral chemical potential. We discuss the solutions of the four cases in which $(T = 0,\mu = 0)$, $(T \neq 0,\mu \neq 0)$, $(T \neq 0,\mu = 0)$ and $(T = 0,\mu \neq 0)$, where the effective potential is so obtained as a function of the background vector field, the chemical potential, and the temperature. The model shows the finite-N corrections generate first-order phase transitions on the self-interacting fermions for the case $N=1$ and the persistence of a second-order phase transition for $N \geq 2$.

\end{abstract}

\maketitle

\section{introduction}
Neutrino physics is one of the keys to understanding the origin of elementary particles beyond the Standard Model (SM).
It is known that neutrinos do not have mass in the SM approach. However, experimental results point out mass
measurements through the transition probabilities in oscillations phenomena \cite{Araki, AthanassopoulosPRL96, AthanassopoulosPRL98}.
The most famous approach to generate masses for the neutrinos is known as see-saw mechanisms \cite{seesaw1,seesaw2, SenjanovicPRD,seesaw3,seesaw4,seesaw5}, that also motivate the models with new particle contents \cite{PatiPRD1974, MohapatraPRD1975, SenjanoviPRD1975, zhang}.
Particularly, neutrinos have three properties combined (mass, chargeless, and spin one-half) which opens the possibility that neutrinos can be Majorana particles \cite{Avignone:2007fu}.

The Lorentz symmetry violation (LSV) also provides another framework beyond the SM in that background $4$-vectors are
introduced and constrained via decay widths, scattering processes, or astrophysical tests \cite{GomesPRD2020, Mewes1}.
This is known as the Standard Model Extension (SME) \cite{Colladay, Colladay2}. In connection with neutrino physics,
an LSV scenario also emerges on effective models with quartic self-interacting couplings involving fermions
in which background fields are introduced as auxiliaries fields that play a fundamental role in the understanding
of the vacuum via effective potential formalism \cite{gomes2022, Assuncao, Assuncao1, Charneski}. The measurements of the neutrino's oscillations impose
exciting constraints on the LSV parameters associated with the vacuum of the fermion quartic models \cite{gomes2022}.
In this sense, fermionic models with quartic self-interactions are also motivated by the study of physics in low dimensions,
as the Thirring model \cite{Thirring}, the Gross-Neveu model \cite{Gross-Neveu}, and the Nambu-Jona-Lanisio (NJL) model for
quantum chromodynamics (QCD) \cite{NJL}. In all these models, the dynamical symmetry breaking shows the vacuum properties
and the possible phase transitions in the presence of a temperature and chemical potential.

Recently, the study of finite-N effects on symmetry breaking and its impact on many systems have been analyzed, as graphene-like systems \cite{kneur1,kneur2, kneur3}, bidimensional semimetals \cite{ygomes23} and nuclear and quark matter as well \cite{kneur4, kneur5}. The non-perturbative calculations have shown new features that pass unseen in the large-N approximations, such as first-order transitions.

Thus, inspired by those intriguing non-perturbative results, in this paper, we investigate the vacuum phase transitions of $N$ massless fermions in the presence of
quartic self-interactions based on the ref. \cite{gomes2022}. The self-interaction between pseudo-current $j^5_\mu=\bar{\psi}\gamma_\mu \gamma_5 \psi$ generates a fermionic condensate that breaks parity (P) and (CPT) symmetries, along with the Lorentz symmetry. In particular, the time component of the pseudo-vector condensate can be interpreted as a chiral chemical potential $\mu_5$, although it is not associated with any conserved charge. Usually, $\mu_5$
is treated as a parameter arising from external dynamics in the slowly varying limit, {\it e.g.}, via effective axion dynamics $\mu_5 \propto \dot{\theta}$, with $\theta$ the axion field. In our case, the chiral chemical potential arises from the dynamical Lorentz symmetry-breaking character of the self-interacting model. A chiral chemical potential generates a difference between the number densities of right- and left-handed chiral fermions, the so-called chiral imbalance. It has been a target of recent studies, and its presence is responsible for several effects, such as the Chiral Magnetic effect \cite{CME1, CME2, CME3}, Chiral vortical effect \cite{CVE1}, Chiral separation effect \cite{CSE1, CSE2}.

Moreover, in the context of the Einstein-Cartan-Kibble-Sciama (ECKS) theory of gravity, as described in \cite{Poplawski:2010kb}, General Relativity is extended to incorporate the matter spin degree of freedom, leading to the generation of axial self-interacting gravitational repulsion in the early Universe. This repulsion prevents the cosmological singularity through repulsive axial self-interactions between fermions. Attractive axial self-interactions can also be generated in a more general spinor dynamics framework where gravity is governed by the Einstein-Cartan-Holst action, as discussed in \cite{Magueijo:2012ug}. Such effective models inspire the present work.

The generating functional approach from quantum field theory
is used to obtain the effective potential in terms of one auxiliary $4$-vector field plus the optimized perturbation theory (OPT) method that brings the non-perturbative results to the model \cite{Fraga09}. The temperature $(T)$ and the chemical potential $(\mu)$ are introduced via Matsubara formalism. Consequently, the Lorentz symmetry is broken in the model, and an effective chiral chemical potential is generated under certain conditions. We chose the time-like $4$-vector for our analysis such that the principle of minimal sensitivity  (PMS) and the gap equations are derivative from the effective potential. Therefore,
we obtain the solutions of these equations for the four cases : (i) $ (T = 0, \mu = 0)$, (ii) $(T \neq 0, \mu \neq 0)$,
(iii) $ (T \neq 0, \mu = 0) $, and (iv) $ (T = 0, \mu \neq 0) $. The solutions for the effective potential are discussed,
and the correspondent vacuum states are shown numerically in cases (i) and (iii) and analytically in the second case (ii).
The paper is organized as follows: In the section (\ref{sec2}), the $N$ 4-component fermions model in the presence of a quartic self-interaction is presented, and the effective potential is obtained in terms of momentum integrals. In the section (\ref{sec3}), the PMS and GAP equations are evaluated in the vacuum states, and the effective potential is studied for the case $(T=0,\mu=0)$ with the subsection (\ref{subsec3}).
We introduce the temperature and chemical potential through the momentum integrals in the section (\ref{sec4}).
This section is subdivided into three subsections: The first subsection (\ref{subsec41}) is for $(T\neq0,\mu=0)$,
and the second one (\ref{subsec42}) is for
$(T = 0,\mu \neq 0)$. The third subsection discusses the phase diagrams (\ref{subsec43}). The section \ref{discuss} presents the discussion of the results and some possible applications. Finally, the conclusions are considered in the section (\ref{sec5}). The useful integrals in the paper
are shown in Appendix A and B.
We use the natural units $\hbar=c=k_{B}=1$, and the Minkowski metric is $\eta^{\mu \nu}= \mbox{diag} (+1,-1,-1,-1)$
for throughout this work.

\section{The $N$ fermions model in the presence of a quartic self-interaction}
\label{sec2}
The following Lagrangian governs the model \cite{gomes2022}:
\begin{equation}\label{lagr1}
\mathcal{L}=
\overline{\psi}_k \, i \, \slashed{\partial} \, \psi_k - \frac{g}{2 N}
\left( \,
\overline{\psi}_{k}\gamma_\mu \gamma_5\psi_{k} \, \right)^2 \; ,
\end{equation}
where $\psi_k$ is a massless four-component fermion, $k=\left\{\,1,...,N\,\right\}$ is the index that specifies the flavors,
the repeated $k$-index means an implicit sum over the flavors, $g$ is a coupling constant with length dimension to the square in
four dimensions,
$\slashed{\partial}=\gamma^{\mu}\partial_{\mu}$ is the slashed partial operator,
and $\gamma^{\mu}$ are the usual Dirac matrices that satisfy the relation $\gamma^\mu \, \gamma^\nu=
\eta^{\mu \nu} \, \mathds{1}-2\,i\,\Sigma^{\mu\nu}$. The auxiliary vector field $A^\mu$ is introduced as follows  :
\begin{equation}\label{lagr2}
\mathcal{L} =
\overline{\psi}_k\left( i \slashed{\partial}- \slashed{A}\gamma_5 \right) \psi_k + \frac{N}{2 g} A_\mu A^\mu \; ,
\end{equation}
where $A^\mu = \frac{g}{N} \, \overline{\psi}_k \gamma^\mu \gamma_5\psi_k$. Therefore, the effective
Lagrangian (\ref{lagr2}) sets the model we investigate with the beyond large-N approach. In previous works \cite{Assuncao, Assuncao1}, the dynamical breaking of Lorentz symmetry on a massless fermionic system has been studied through the model represented by equation \eqref{lagr2} by integrating out the fermionic fields and assuming the large-N limit to compute the effective potential.
%

%

\begin{figure}[h]
\includegraphics[scale=0.37]{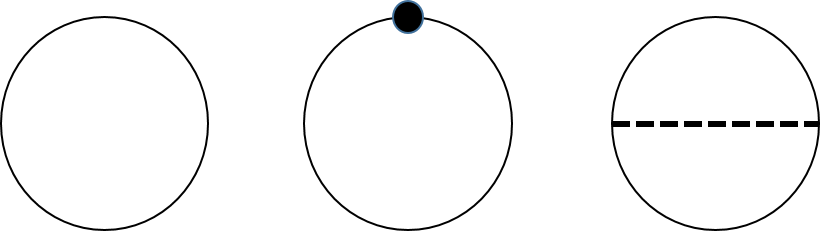}
\vspace{0.2cm}
\caption{The Feynman diagrams contributing up to linear $\delta$-order
to the effective potential. The first sets the effective potential in the large-N limit. The black
dot represents a $\delta$-insertion. The last diagram brings the first correction to the effective potential
due to the self-energy (figure based on ref. \cite{kneur1}).}
\label{feyn}
\end{figure}

%
In order, the introduction of
corrections beyond the large-N limit to the model are implemented via the optimized perturbation theory (OPT)
deformation based on \cite{kneur1, kneur2,oko87, yukalov}, through the fictitious $\delta$-parameter :
%
%
%
\begin{eqnarray}\label{Ldelta}
\mathcal{L}_\delta &=& (1- \delta)\mathcal{L}_0(\mathtt{a}) + \delta \mathcal{L}=
\nonumber \\
&&\hspace{-1.cm}
=\overline{\psi}_k\left[ \, i \slashed{\partial} - \left(\slashed{\mathtt{a}} - \left(\slashed{\mathtt{a}}-\slashed{A} \right)\delta \right)\gamma_5 \, \right] \psi_k +  \frac{\delta N}{2 g} A_\mu A^\mu \; =
\nonumber \\
&&\hspace{-1.cm}
=\overline{\psi}_k\left( \, i \slashed{\partial} -\slashed{\mathtt{a}}'\gamma_5 \, \right) \psi_k +  \frac{\delta N}{2 g} \, A_\mu A^\mu \; ,
\end{eqnarray}
%
where $\mathtt{a}^{\mu}$ is a regulator $4$-vector field containing the finite-N corrections information,
with $\slashed{\mathtt{a}}'= \slashed{\mathtt{a}} - \left(\slashed{\mathtt{a}}-\slashed{A} \right)\delta$, and $\mathcal{L}_0$ is the kinetic term of the fermionic Lagrangian coupled to $\mathtt{a}^\mu$. The correspondent classical motion equations for the auxiliary fields $A^{\mu}$ and $\mathtt{a}^\mu$ are, respectively, read as
\begin{subequations}
\begin{eqnarray}
\frac{\delta N}{g} \, A_\mu - \delta \left( \overline{\psi}_k \gamma_\mu \gamma_5 \psi_k \right) =0 \; ,
\label{EqA}
\\
(\delta-1) \, \overline{\psi}_k \gamma_\mu \gamma_5 \psi_k =0 \; ,
\label{Eqa}
\end{eqnarray}
\end{subequations}
in which the eq. (\ref{Eqa}) implies that $\delta = 1$, since that $\overline{\psi}_k \gamma_\mu \gamma_5 \psi_k \neq 0$,
and we recover the original lagrangian of \eqref{lagr2}, where $\mathtt{a}^{\mu}$ disappears naturally in (\ref{Ldelta}).
As usual in quantum field theory, one can integrate the fermionic degree of freedom, resulting in an effective action
for the bosonic field $A^{\mu}$. The perturbative approach defines the generating functional associated with (\ref{Ldelta}) as
\begin{equation}
Z_\delta = \int {\cal D}A^\mu \, {\cal D}\overline{\psi} \, {\cal D}\psi \, e^{i \, \int d^4x \,
{\cal L}_{\delta} } = \int {\cal D}A^\mu \, e^{i S_{eff}[A,\delta]} \; ,
\end{equation}
where the effective action is
\begin{eqnarray}
S_{eff}[A,a,\delta]
=- N \int d^{4}x \, V_{eff}(A,a,\delta) \; .
\end{eqnarray}
%
%
%
The effective potential expanded up to linear order on the $\delta$-parameter is given by
\begin{eqnarray}\label{VeffInt}
&&
\,V_{eff}(A^\mu,\mathtt{a}^{\mu},\delta) = -\frac{\delta}{2 g} \, A_\mu A^\mu
\nonumber \\
&&
+i \int \frac{d^4p}{(2\pi)^4} \, \mbox{Tr} \ln \left( \slashed{p} - {\slashed{\mathtt{a}}}\gamma_5 \right)
\nonumber \\
&&
+ i  \delta \int \frac{d^4p}{(2\pi)^4} \, \mbox{Tr} \left[\frac{ (\slashed{\mathtt{a}} - \slashed{A})\gamma_5}{\slashed{p} - \slashed{\mathtt{a}} \gamma_5}\right]   \nonumber \\
&&
+ \frac{i}{2}  \int \frac{d^4p}{(2\pi)^4} \, \mbox{Tr} \Bigg[ \, \frac{\Sigma(\mathtt{a}) }{\slashed{p} - \slashed{\mathtt{a}}\gamma_5} \, \Bigg] + O(\delta^2) \; .
\hspace{1.0cm}
\end{eqnarray}
The numerator of the third trace in (\ref{VeffInt}) is
\begin{equation}\label{Sigmaa}
\Sigma(\mathtt{a}) \approx  \frac{\delta g}{ N} \int \frac{d^4q }{(2\pi)^4}\left[ \gamma_\mu \gamma_5\frac{i}{\slashed{q} - \slashed{\mathtt{a}}\gamma_5 }\gamma^\mu \gamma_5 \right] + O(\delta^2) \, ,
\end{equation}
that also is expanded to the first order of the $\delta$-parameter, and the trace (Tr) acts on the spinor space.
{Therefore, in this stage of calculations, one applies a truncation on the $\delta$-expansion up to $\delta$
and our results will show the truncation is sufficient to extract finite-N corrections.}
The correspondent diagrams of these contributions are illustrated in the figure (\ref{feyn}).
\vspace{0.5cm}
The common fraction of the integrals in (\ref{VeffInt}) can be written as
\begin{equation}
\frac{1}{\slashed{p}-\slashed{\mathtt{a}}\gamma_5 }= \frac{N_0(p)}{R(p)} \; ,
\end{equation}
in which the numerator is
$N_0(p) = (p^2+  \mathtt{a}^2 +2  p \cdot \mathtt{a} \gamma_5 )( \slashed{p} + \slashed{\mathtt{a}} \gamma_5)$,
and the denominator is $R(p) = (p+\mathtt{a})^2(p-\mathtt{a})^2$.
Thereby, the regularized effective potential can be written as
\begin{widetext}
\begin{eqnarray}\label{Veff}
\,V_{eff}(A^\mu,\mathtt{a}^{\mu},\delta,D) = -\frac{\delta}{2 g} A_\mu A^\mu +  \, \Omega(\mathtt{a},D) + \delta \left(\mathtt{a}_{\mu} - A_{\mu} \right)
\, \Pi^{\mu}(\mathtt{a},D)  - \frac{\delta g}{2 N} \, \Xi(\mathtt{a},D) \; ,
\end{eqnarray}
\end{widetext}
where $\Omega(\mathtt{a},D)$, $\Pi^\mu(\mathtt{a},D)$ and $\Xi(\mathtt{a},D)$ are defined by the regularized integrals
\begin{eqnarray}
\Omega(\mathtt{a},D) =
i\,\Lambda^{4-D} \!\! \int \frac{d^D\bar{p}}{(2\pi)^D} \, \mbox{Tr}  \ln \left( \slashed{\bar{p}}- \slashed{\mathtt{a}}\gamma_5\right) \; , \;\;
\label{intOmega}
\end{eqnarray}
\begin{eqnarray}
\Pi^\mu(\mathtt{a},D) = i\,\Lambda^{4-D} \!\! \int \frac{d^D\bar{p}}{(2 \pi)^D} \, \mbox{Tr} \left[ \frac{\gamma^\mu \, \gamma_5}{\slashed{\bar{p}}- \slashed{\mathtt{a}}\gamma_5} \right] \; , \;\;
\label{intPi}
\end{eqnarray}
\begin{eqnarray}
\Xi(\mathtt{a},D) \!\!&=&\!\! \Lambda^{8-2D} \!\! \int \frac{d^D\bar{p}}{(2\pi)^D}
\frac{d^D\bar{q}}{(2\pi)^D}
\nonumber \\
&&
\times \,
\frac{\mbox{Tr}\left[ \,N_0(\bar{p})\gamma_\mu \gamma_5\,N_0(\bar{q})\gamma^\mu \gamma_5\,\right]}{R(\bar{p})\,R(\bar{q})} \; ,
\hspace{0.7cm}
\label{intXi}
\end{eqnarray}
%
%
%
%
%
%
We also have introduced the dimensional regularization $(D)$ substituting the divergent momentum integrals from
(\ref{VeffInt}) and (\ref{Sigmaa}) by $\int d^{4}p/(2\pi)^4 \rightarrow \Lambda^{4-D} \, \int d^{D} \bar{p}/(2\pi)^D$,
where $\Lambda$ is a parameter with mass dimension, $\bar{p}$ and $\bar{q}$ are the $4$-momentum defined in $D$-dimensions.
The modified integrals are finite and naturally diverge in the ultraviolet regime when $D=4$. Consequently,
the dimensional parameter $D$ also regularizes the effective potential, and the Ward identity is not violated in this regularization scheme.
{ Using the same techniques from the refs. \cite{Assuncao,Mariz}, the regularized integrals yield the results in the limit $D \rightarrow 4$
(see the Appendix A for details):
\begin{subequations}
\begin{eqnarray}
\Omega(\mathtt{a}) \!&=&\! \frac{(\mathtt{a}\cdot\mathtt{a})^2}{12\pi^2} + \text{constant}\,  ~~,
\label{partialOmegaD4}
\\
\Pi^{\mu}(\mathtt{a}) \!&=&\! -\frac{\mathtt{a}\cdot\mathtt{a}}{3\pi^2} \, \mathtt{a}^{\mu} \; ,
\label{PimuD4}
\\
\Xi(\mathtt{a}) \!&=&\! \frac{(\mathtt{a}\cdot\mathtt{a})^3}{18\pi^4} \; .
\label{XiD4}
\end{eqnarray}
\end{subequations}
}

In the next section, we investigate the contributions of the finite parts for the effective potential in the case of the time-like background fields.
%

%


%
\section{PMS and gap equation beyond the N-large limit}
\label{sec3}
The principle of minimal sensitivity (PMS) applied to the effective potential (\ref{Veff}) is defined by the following identity:
\begin{equation}\label{EqVeffa}
\frac{\partial V_{eff}}{\partial \mathtt{a}_\mu}\Bigg{|}_{\delta=1} = 0 \; ,
\end{equation}
that yields the PMS equation
\begin{widetext}
\begin{eqnarray}\label{PMSeq}
\left[ \,  \, \frac{\partial \Omega}{\partial \mathtt{a}_\mu} +\Pi^\mu(\mathtt{a},D)+
\left(\mathtt{a} - A\right)^\nu \, \frac{\partial \Pi_\nu}{\partial \mathtt{a}_\mu}
- \frac{ g}{ 2 N}  \frac{\partial \Xi}{\partial \mathtt{a}_\mu} \, \right] \Bigg{|}_{\delta=1}\!\!\!=0 \; .
\end{eqnarray}
\end{widetext}
The solutions of equation \eqref{PMSeq} given by $\mathtt{a}^{\mu}$ depend on various original parameters
including the couplings and the pseudo-vector field $A^{\mu}$. These results in non-perturbative dependencies
in the model's coupling and other parameters. The gap equation is defined by
\begin{equation}
\frac{\partial V_{eff}}{\partial A_\mu}\Bigg{|}_{A^{\mu}=\bar{A}^{\mu},\mathtt{a}^{\mu}=\bar{\mathtt{a}}^{\mu},\delta=1} = 0 \; ,
\end{equation}
that yields the relation
\begin{equation}\label{Gapeq}
\bar{A}^\mu +  g\, \Pi^\mu (\bar{\mathtt{a}},D) = 0 \; .
\end{equation}
{The equation (\ref{Gapeq}) is evaluated at the minimum states
that we denote as the background $4$-vector and regulator $4$-vector fields, $\bar{A}^{\mu}$ and $\bar{\mathtt{a}}^{\mu}$, respectively.
The $A^{\mu}$ means a state of the model, whereas $\mathtt{a}^{\mu}$ is an auxiliary state that is function of $\bar{A}^{\mu}$ by the equation (\ref{PMSeq}). In the next subsection, we investigate the vacuum structure of the model, {\it i.e.}, the particular case in which the temperature
and the chemical potential are null.}

%

%
\subsection{The case $T=0$ and $\mu = 0$}
\label{subsec3}
For our analysis of the phase transitions, we consider time-like background fields such that $A^{\mu}=(A^{0},{\bf 0})$ 
and $\mathtt{a}^{\mu}=(\mathtt{a}^{0},{\bf 0})$ in the integrals (\ref{intOmega})-(\ref{intXi}), and in the effective potential (\ref{Veff}). 
Under these assumptions, the physical results of the integrals (\ref{intOmega})-(\ref{intXi}) in the limit $D \rightarrow 4$ are reduced to
{
\begin{subequations}
\begin{eqnarray}
\Omega(\mathtt{a}_0) \!&=&\! \frac{\mathtt{a}_0^4}{12\pi^2} 
\; ,
\\
\Pi^{0}(\mathtt{a}_0) \!&=&\! -\frac{\mathtt{a}_0^3}{3\pi^2} \; ,
\\
\Xi(\mathtt{a}_0) \!&=&\!\frac{\mathtt{a}_0^6}{18 \pi^{4}} \; ,
\;\;\;\;\;\;
\end{eqnarray}
\end{subequations}
}

Therefore, the effective potential in this case is
{
\begin{eqnarray}\label{Veffa0A0}
\,V_{eff}(A_0,\mathtt{a}_0) \!&=&\!
-\frac{A_0^2}{2 g}-\frac{\mathtt{a}_0^4}{4\pi^2}
+\frac{\mathtt{a}_0^3 \, A_0}{3 \pi^2}+
\nonumber \\
&&\hspace{-1.cm}
-\frac{g\,a_{0}^6}{36 \pi^4 N}
\; . \;\;\;\;\;
\end{eqnarray}
}
The PMS and GAP equations are, respectively, given by
{
\begin{subequations}
\begin{eqnarray}
A_0 \!&=&\! \frac{ g \, \mathtt{a}_0^3}{6 \pi^2 N}+\mathtt{a}_0 \; ,
\\
\bar{A}_0 \!&=&\! g \, \frac{\bar{\mathtt{a}}_0^3}{3\pi^2}  \; ,
\end{eqnarray}
\end{subequations}
}
in which the vacuum state is set by the time-like vector $\bar{A}^{\mu}=(\bar{A}^0,{\bf 0})$, and
$\bar{\mathtt{a}}^{\mu}=(\bar{\mathtt{a}}^0,{\bf 0})$ is the correspondent auxiliary state.
The asymptotic behavior of the PMS solution is the $N$-large limit (LN), when $N \rightarrow \infty$,
is reduced to :
{
\begin{eqnarray}
\mbox{LN} \; : \; \mathtt{a}_0 = A_0
\hspace{0.3cm} , \hspace{0.3cm}
\bar{A}_0^2-\frac{3\pi^2}{g} =0 \; .
\end{eqnarray}
}
Thereby, the large-$N$ limit the symmetry is broken, if $g=|g|$, and the gap solution is given by $\bar{A}_0 = \frac{\sqrt3 \pi}{\sqrt{|g|}}$, that confirms the result obtained in the ref. \cite{Assuncao}. In the limit $N \rightarrow \infty$, the effective potential is
{
\begin{equation}
V_{eff}^{LN}(A_0)= -\frac{A_0^2}{2 g}+\frac{A_0^4}{12 \pi^2} \; .
\end{equation}
}

\begin{figure}[h]
\includegraphics[scale=0.3]{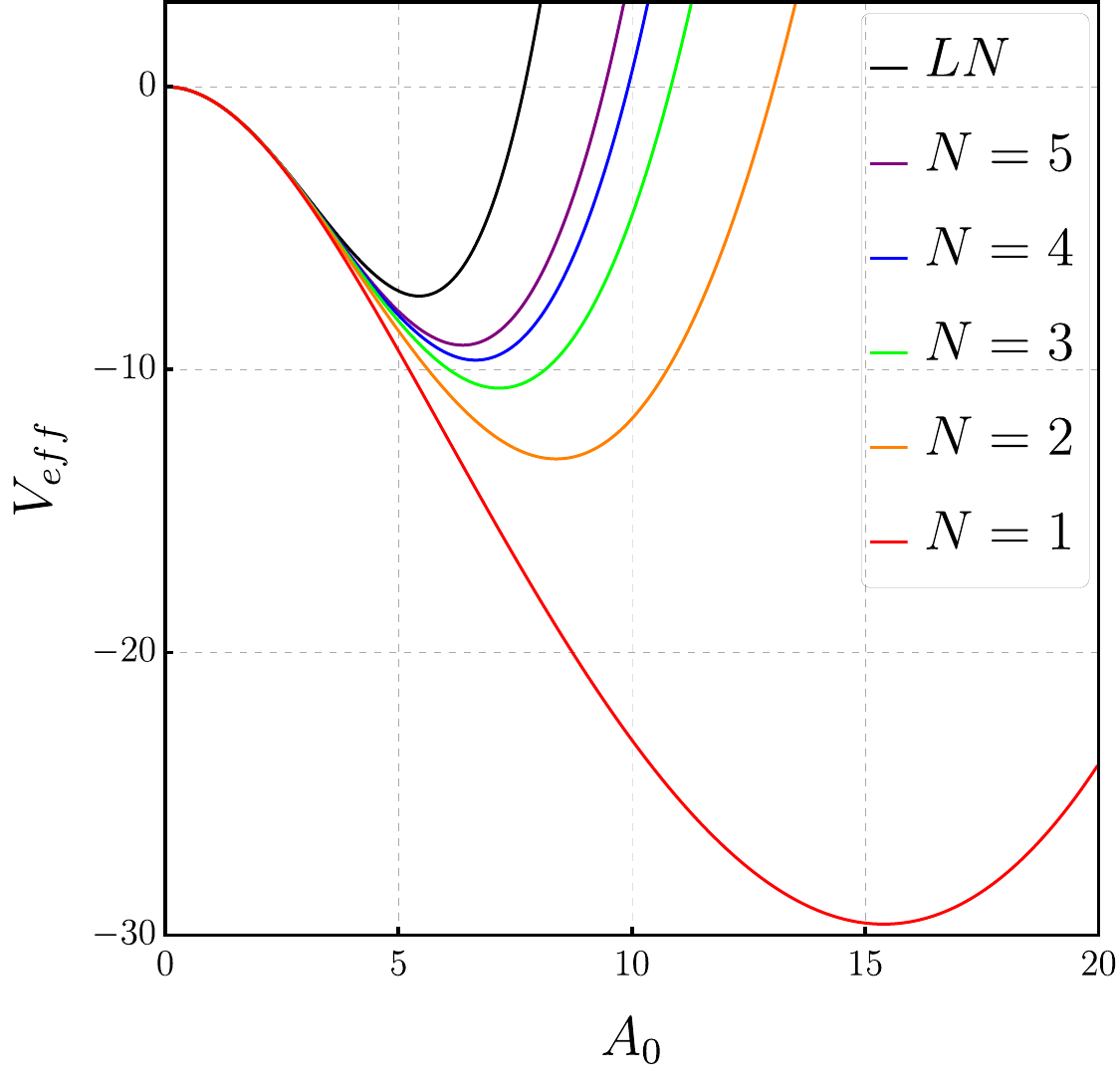}
\caption{The plot of the effective potential $(V_{eff})$ as function of the variable $A_0$, in units of $g^{-1/2}$. In this figure,
we choose $N=\left\{\,1,2,3,4,5\,\right\}$ (colored lines). 
The black line sets the limit $N \rightarrow \infty$.}
\label{fig1}
\end{figure}

Clearly, the nontrivial minimum emerges when $g>0$. Although the analytic form of the finite-N contributions to the effective potential has an intricate non-linear character, one can plot it via numerical implementation. The N-finite effective potential $V_{eff}$ is plotted in Fig. \eqref{fig1} as a function of the variable $A_0$. The black line means the case of $N$-large limit, and we choose $N=\left\{\,2,3,4,5\,\right\}$ represented by the colored lines, in units of $g^{-1/2}$ (that means in energy units). For $N>1/2$, the gap equation reads the solution :
{
\begin{equation}\label{gap0sol}
\bar{A}_0= \frac{\sqrt3 \, \pi }{\sqrt{|g|} } \frac{1}{\left(1-\frac{1}{2N}\right)^{3/2}} \; .
\end{equation}
}
It is worth highlighting that this extra factor reflects the non-perturbative nature of the OPT approach.
%

%
%
Next, we introduce the temperature $(T)$ and chemical potential $(\mu)$ via Matsubara formalism modifying the $p^{0}$-component by $p^{0} \rightarrow i\omega_{n}-\mu$, and summing over all the Matsubara frequencies $\omega_{n}$. This modification is clarified in Appendix B. Thereby, the formalism opens the three possible cases $(T \neq 0 , \mu \neq 0)$, $(T \neq 0 , \mu = 0)$, and $(T = 0 , \mu \neq 0)$ that will be discussed in the next subsections.
%

%
\section{Introducing the temperature and chemical potential}
\label{sec4}
Whenever the thermodynamic quantities $T$ and $\mu$ are present, the translational symmetry in time is broken but kept for the space coordinates.
Thereby, we consider the $\mathtt{a}^{\mu}$ as a time-like $4$-vector $\mathtt{a}^{\mu} = (\mathtt{a}^0,{\bf 0})$, without any loss of generality. 
This choice is also justified in the case of a non-null temperature in which the phase transition analysis emerges from a time-like background vector, see \cite{Assuncao}.
The $\Omega$-integral (\ref{intOmega}) in the presence of a temperature $(T)$, and of a chemical potential $(\mu)$ is so written as :
\begin{eqnarray}\label{Omegaa0}
\Omega(\mathtt{a}_0)  \!&=&\! \frac{T^4}{\pi^2} \! \sum_{\lambda=-1}^{+1}\left[  \text{Li}_4\left(-e^{- \beta\mu_\lambda }\right)
+ \text{Li}_4\left(-e^{  \beta\mu_\lambda }\right) \right] \; ,
\nonumber \\
&&
\end{eqnarray}
where $\beta=T^{-1}$ is the inverse of the temperature, $\mu_\lambda = \mu - \lambda \, \mathtt{a}_0 \, (\lambda=\pm1)$ is the chemical potential summed to the time-component $\mathtt{a}_0$, and $\text{Li}_4$ is poly-logarithm function of fourth degree. The integral (\ref{intPi}) for $T \neq 0$ is

\begin{eqnarray}
\Pi^{0}(\mathtt{a}_0)  =  Y(\mu_-) - Y(\mu_+) \; ,
\end{eqnarray}
where
\begin{eqnarray}\label{f2}\nonumber
Y(\mu_\pm) \!&=&\!
\int \frac{d^{3}p}{(2 \pi)^{3}} \left[ \frac{1}{1+e^{\beta(|{\bf p}| + \mu_\pm)}} - \frac{1}{1+e^{\beta(|{\bf p}| - \mu_\pm)}}\right]
\nonumber \\
&&
\hspace{-1cm}
=\frac{4T^3}{\pi^2} \left[ \, \text{Li}_3\left(-e^{\beta\mu_\pm }\right)- \text{Li}_3\left(-e^{-\beta\mu_\pm }\right) \, \right] \; ,
\;\;\;
\end{eqnarray}
in which $\text{Li}_{3}$ is a Poly-logarithm function of third degree. The $\Xi$-integral (\ref{intXi}), for $T \neq 0$, is read
\begin{eqnarray}
\Xi(\mathtt{a}_0,D) =  Y(\mu_-)^2+ Y(\mu_+)^2 \; ,
\end{eqnarray}
Using these results, we can discuss
the particular cases of $(T \neq 0 , \mu = 0)$ and $(T = 0 , \mu \neq 0)$ in the next subsection.

\subsection{The case $T \neq 0$ and $\mu = 0$}
\label{subsec41}
\begin{figure}[h]
\includegraphics[scale=0.32]{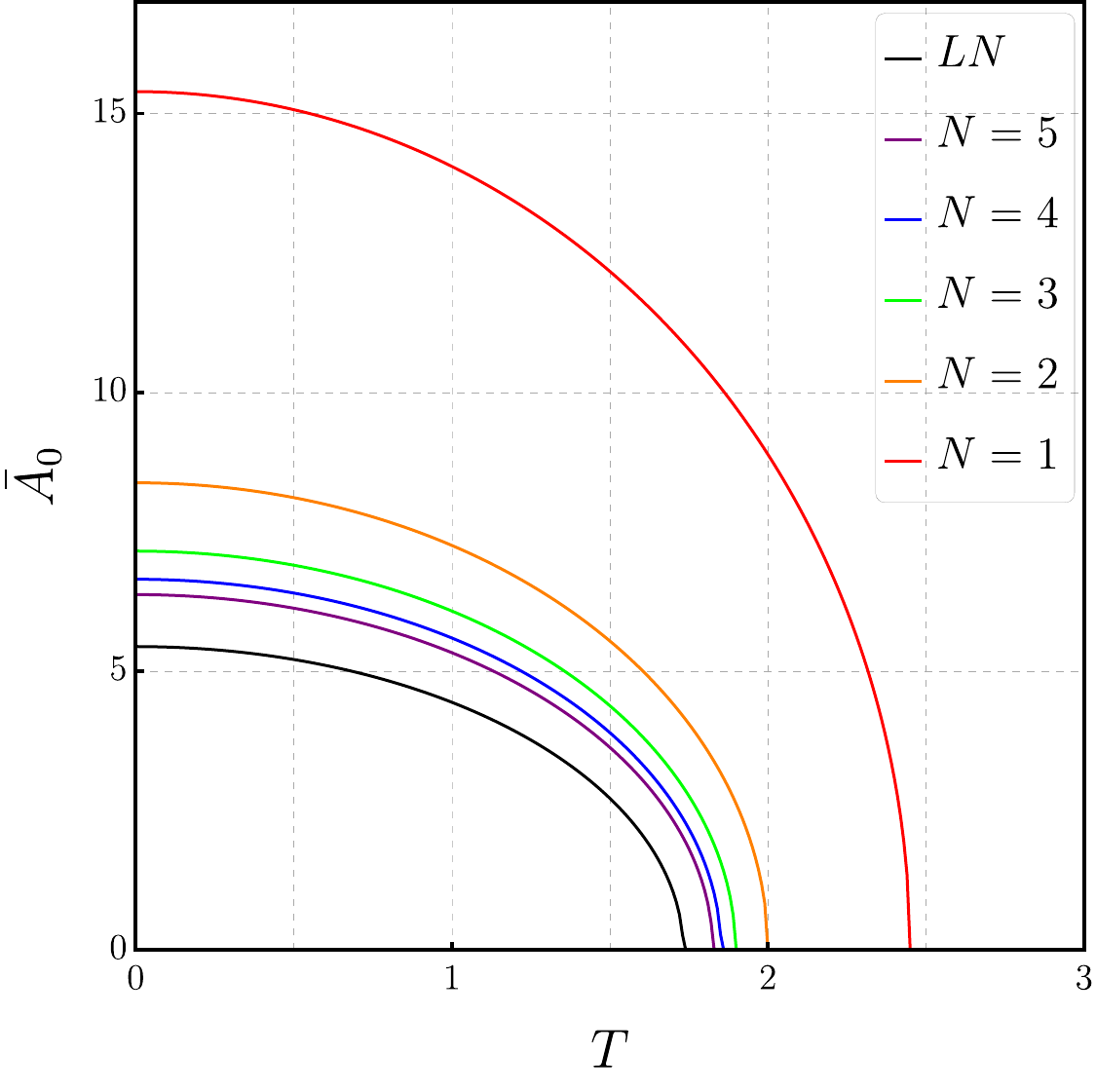}
\caption{The plot of the gap solution as function of the temperature $T$ at $\mu=0$, in units of $g^{-1/2}$. In this figure,
we choose $N=\{1,2,3,4,5\}$ (colored lines). The black line sets the limit $N \rightarrow \infty$.}
\label{gap2}
\end{figure}

Taking the limit $\mu \rightarrow 0$, the gap equation reads :
\begin{equation}
\bar{A}_0 = \frac{2 g T^3}{\pi^2} \left[\text{Li}_3\left(-e^{-\frac{\bar{\mathtt{a}}_0}{T}}\right)-\text{Li}_3\left(-e^{\frac{\bar{\mathtt{a}}_0}{T}}\right)\right] \; .
\end{equation}
The PMS equation in the limit $\mu \rightarrow 0$ is
\begin{eqnarray}\nonumber
&& \left[\text{Li}_2\left(-e^{\mathtt{a}_0/T}\right)+\text{Li}_2\left(-e^{-\mathtt{a}_0/T}\right)\right] \times
\\
\nonumber
&&
\left[ \mathtt{a}_0-A_0-\frac{ g \, T^3}{\pi ^2 N} \, \text{Li}_3\left(-e^{\mathtt{a}_0/T}\right)
+\right. \\
&&
\left. + \frac{ g \, T^3}{\pi ^2 N} \, \text{Li}_3\left(-e^{-\mathtt{a}_0/T}\right)\right]=0 \; ,
\end{eqnarray}
in which we can check that $\mathtt{a}_0=0$ implies into $A_0 = 0$. Combining both equations, one can find the numerical solution for $\bar{A}_0(T)$. The solution is illustrated in Fig. \ref{gap2} for representative values of $N$ and $g$. Substituting the gap equation in the PMS equation and taking the limit $\bar{\mathtt{a}}_0 \to 0$, we obtain the solution for the critical temperature
%
\begin{equation}
T_c =\frac{\sqrt{3}}{\sqrt{g}}\frac{1}{\sqrt{1-1/2N}} \;, N\geq 1 .
\end{equation}
%
The $N$-large limit fixes the critical temperature at $T_{c}=\frac{\sqrt3}{\sqrt{g}}$.

\subsection{The case $T=0$ and $\mu \neq 0$}
\label{subsec42}
In the case of $T \rightarrow 0$, we need to use the limits:
\begin{subequations}
\begin{eqnarray}
\lim_{T \rightarrow 0}\left[\frac{1}{1+ e^{\frac{|{\bf p}| \pm \mu_\lambda}{T}}} \right] \!\!&=&\!\! \Theta(-|{\bf p}| \mp \mu_\lambda)
\, , \;\;\;\;
\label{limit1}
\\
\lim_{T \rightarrow 0} T \ln \left[1+ e^{-\frac{|{\bf p}|+ \mu_\lambda}{T}} \right] \!\!&=&\!\! -\left(|{\bf p}|+ \mu_\lambda\right)
\nonumber \\
\times \, \Theta(-|{\bf p}|- \mu_\lambda) \; ,
\label{limit2}
\\
\lim_{T \rightarrow 0} T \ln \left[1+ e^{-\frac{|{\bf p}|- \mu_\lambda}{T}} \right] \!\!&=&\!\! \left(-|{\bf p}|+ \mu_\lambda\right)
\nonumber \\
\times \, \Theta(-|{\bf p}|+ \mu_\lambda) \; ,
\label{limit3}
\end{eqnarray}
\end{subequations}
in which the $\Theta$-Heaviside function satisfies the conditions $\Theta(x)=0$ for $x<0$, and $\Theta(x)=1$ for $x>0$.
Using these properties, the $\Omega$-integral is
\begin{equation}
\lim_{T \rightarrow 0}\Omega(\bar{\mathtt{a}}_0)
=\frac{1}{12\pi^2}(6 \, \bar{\mathtt{a}}_0^2 \, \mu^2+\bar{\mathtt{a}}_0^4+\mu^4) \; .
\end{equation}
Analogously, the result for $\Pi^{\mu}$ is

\begin{eqnarray}
\lim_{T \rightarrow 0}\Pi^{0}(\bar{\mathtt{a}}_0)
= - \, \frac{\bar{\mathtt{a}}_0}{3 \pi ^2}\,(\bar{\mathtt{a}}_0^2+3\mu ^2) \; .
\end{eqnarray}

\begin{figure}[h]
\includegraphics[scale=0.35]{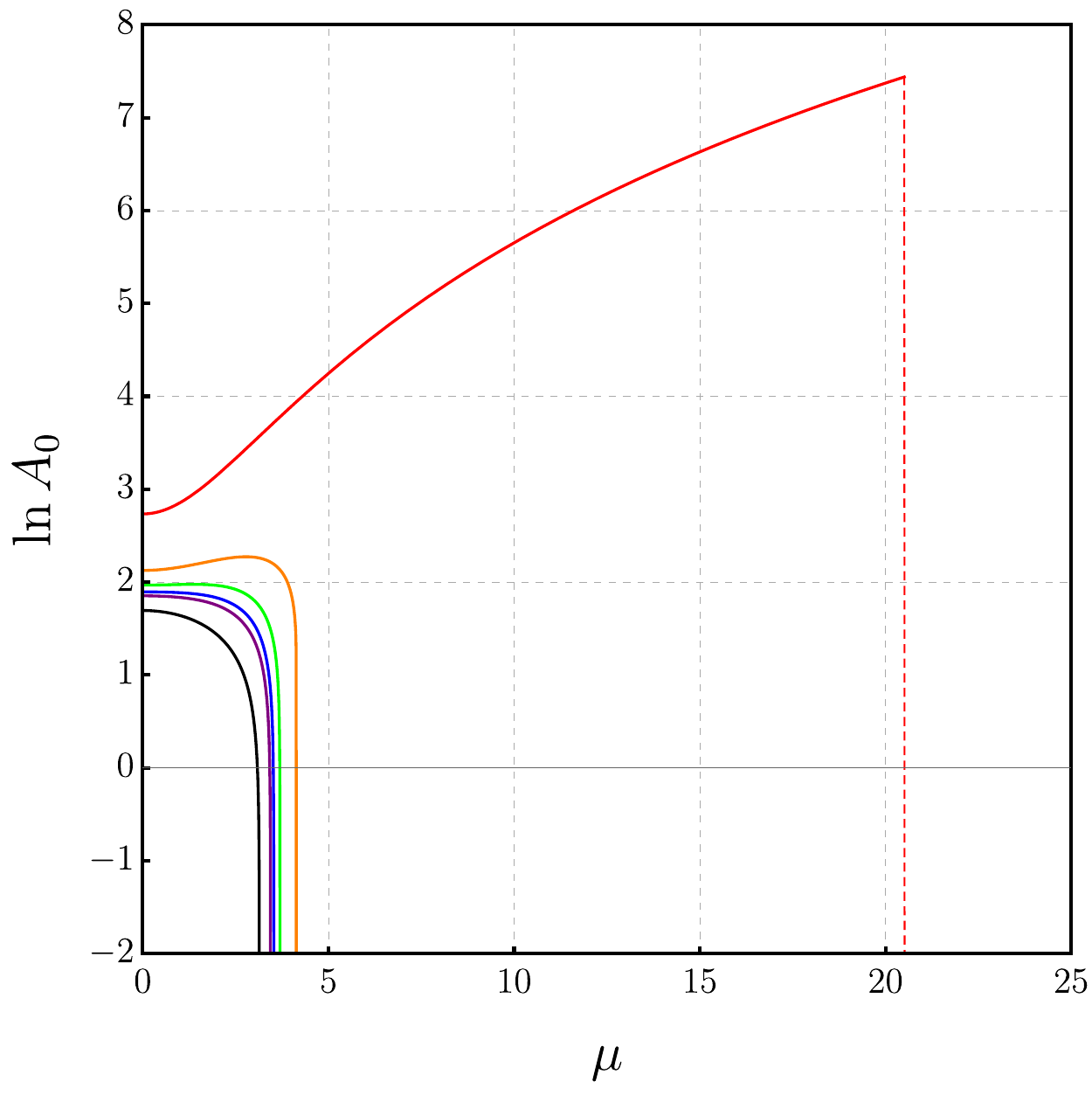}
\caption{The logarithm plot of the gap solution as a function of the chemical potential $(\mu)$, when $T=0$, in units of $g^{-1/2}$. In this figure,
we choose $N=\{1,2,3,4,5\}$ (colored lines) and $g=1$ (in energy units). The black line represents the $N$-large limit. 
The red dashed line indicates the discontinuity for $N=1$ at $\mu\approx 20.51$. }
\label{fig4}
\end{figure}

The $\Xi$-integral is reduced to
\begin{eqnarray}\nonumber
\lim_{T \rightarrow 0} \Xi(\bar{\mathtt{a}}_0)&=&\frac{(\bar{\mathtt{a}}_0-\mu)^6}{36 \pi^4}+\frac{(\bar{\mathtt{a}}_0+\mu)^6}{36\pi^4}
\\
&&
\hspace{-1.5cm}
= \frac{1}{18\pi^4}(\bar{\mathtt{a}}_0^6 + 15 \bar{\mathtt{a}}_0^4 \, \mu^2 + 15 \bar{\mathtt{a}}_0^2 \, \mu^4 + \mu^6) \; .
\;\;\;\;
\end{eqnarray}
Thereby, the PMS and GAP equations are, respectively, given by
\begin{subequations}
\begin{eqnarray}\nonumber
&&
(\mathtt{a}_0-A_0) \left(\mathtt{a}_0^2+\mu^2\right) + \\
&&+\frac{\mathtt{a}_0 g}{6 \pi ^2 N} \left(10 \mathtt{a}_0^2 \mu^2+\mathtt{a}_0^4+5 \mu^4\right) = 0 \; ,
\\
&&
-\frac{\bar{A}_0}{g}+\frac{\bar{\mathtt{a}}_0}{3 \pi ^2} \, (\bar{\mathtt{a}}_0^2+3 \mu^2) = 0 \; .
\hspace{1.0cm}
\end{eqnarray}
\end{subequations}
{ By combining both equations, one reaches the following identity:
\begin{eqnarray}\nonumber
    &&2 \bar{\mathtt{a}}_0^2 \left[g \mu^2 (4 N-5)-3 \pi^2 N\right]
    +\\
    \nonumber
    &&+\bar{\mathtt{a}}_0^4 g (2N-1)+g \mu^4 (6 N-5)-6 \pi ^2 \mu^2 N =0 ~~,\\
\end{eqnarray}
The above equation can be combined, and the solutions are given by:
\begin{equation}
    \bar{\mathtt{a}}_0^2 = \frac{g \mu^2 (5-4 N)+3 \pi ^2 N\pm \sqrt{\Delta }}{g (2 N-1)}~~,
\end{equation}
with
\begin{eqnarray}\nonumber
    && \Delta = 4 g^2 m^4 (N-5) (N-1)-12 \pi ^2 g \mu^2 (N-2) N\\
    &&+9 \pi ^4 N^2~~.
\end{eqnarray}
}
{Through the gap equation, one can find the solution for $\bar{A}_0$ as a function of $\mu$. This is shown in Fig. \ref{fig4} for some representative values of $N$ and $g$. Once again, we can check that $\bar{\mathtt{a}}_0=0$ implies into $\bar{A}_0 = 0$.  In the limit $N \rightarrow \infty$, the solution to the PMS equation reaches $\mathtt{a}_0=A_0$, and using this result, one finds the solution to the gap equation :
\begin{equation}
\mbox{LN} :  \bar{A}_0 = \frac{\sqrt{3}\,\pi}{\sqrt{g}} \, \sqrt{1-\frac{g \, \mu^2}{\pi^2}} \; ,
\end{equation}
and the critical chemical potential is given by $\mu_c^{LN} = \frac{\pi}{\sqrt{g}}$. Interestingly, the solution of the GAP and PMS equations for $N=1$ is :
\begin{equation}
\bar{\mathtt{a}}_0^2= \mu^2+\frac{3\pi^2}{g}+\frac{3\pi^2}{g} \sqrt{1+\frac{4 g \mu^2}{3 \pi^2} } \; ,
\end{equation}
which yields a real solution for any value of $\mu$. The result differs significantly from the results for $N>1$, where there is always some critical chemical potential in which the solution becomes imaginary. This feature shows that the $N=1$ case is peculiar, in the sense that, in low temperatures, one reaches a first-order phase transition, as is shown in Fig. \ref{fig4}.  }
Using the analysis of the PMS equation in the limit $\bar{\mathtt{a}}_0 \to 0$, after the application of the gap equation, the critical value of chemical potential $(\mu_{c})$ satisfies the equation
\begin{equation}\label{eqmuc}
\mu^2_c \, g \, (6N-5) - 6 \, N \, \pi^2 =0 \; ,
\end{equation}
where the symmetry is restored as the solution of (\ref{eqmuc}). The solution is given by
%
%
\begin{equation}
\mu_c = \frac{\pi}{\sqrt{g}}\frac{1}{\sqrt{1-\frac{5}{6N}}} \; ,  N>1  ~,
\end{equation}
%
in which the critical chemical potential is constrained by the condition $g > 0$. { For $N=1$, the value of the critical chemical potential can be obtained numerically and is given by $\mu_c(N=1) \approx 20.51$ (in units of $g^{-1/2}$).}

%


%

\subsection{Phase portrait}
\label{subsec43}

The phase diagram is shown in the fig. \eqref{fig6}, and one can see that the critical curve is modified in the finite-$N$ approach. In the cases where $N\geq2$, besides the quantitative modification, there is no qualitative difference, {\it i.e.},  the phase transition still has a continuous behavior and thus is a second-order phase transition. On the other hand, the case $N=1$ shows a qualitative difference. Thus, to access the information about the density, one needs to start from the thermodynamic potential defined as  $\Phi= V_{eff}(\bar{A}_0,\bar{\mathtt{a}}_0)$. Now, the density can be properly and consistently defined as
\begin{equation}
-\frac{\partial\Phi}{\partial \mu}  = n -  \frac{\partial \bar{A}_0}{\partial \mu}\frac{\partial\Phi}{\partial \bar{A}_0} - \frac{\partial \bar{\mathtt{a}}_0}{\partial \mu}\frac{\partial\Phi}{\partial \bar{\mathtt{a}}_0} \; ,
\end{equation}
and using the PMS and gap equations, one shows that
\begin{eqnarray}
n = -\frac{\partial\Phi}{\partial \mu} \; .
\end{eqnarray}
This expression is used to construct the phase portrait in Figs. \ref{fig6} and \ref{fig8}. Table \ref{tab:my_label} shows some representative values of critical temperature and chemical potential.
\begin{table}[H]
    \centering
    \begin{tabular}{|l|l|l|l}
    \hline ~~N~~ & ~~ $T_c(\mu=0)$ ~~&~~ $\mu_c (T=0)$ ~~ \\
      \cline{1-3} ~~ $\infty$ ~~ & ~~$1.73$ ~~&~~ $3.14$~~ \\ \cline{1-3}
        ~~$1$~~ & ~~$2.45$ ~~&~ $20.51$~~ \\ \cline{1-3}
     ~~$2$~~ & ~~$2.00$~~ & ~~$4.11$~~\\ \cline{1-3}
      ~~$3$~~ & ~~$1.89$~~ & ~~$3.69$~~\\ \cline{1-3}
     ~~$4$~~ & ~~$1.85$~~ & ~~$3.53$~~ \\ \cline{1-3}
     ~~$5$~~ & ~~$1.82$~~ &~~$3.44$~~ \\ \cline{1-3}
    \end{tabular}
    \caption{Table of the values of the critical temperature $T_c$ at $\mu=0$ and critical chemical potential $\mu_c$ at $T=0$; in units of $g^{-1/2}$.}
    \label{tab:my_label}
\end{table}

\begin{figure}[H]
\includegraphics[scale=0.3]{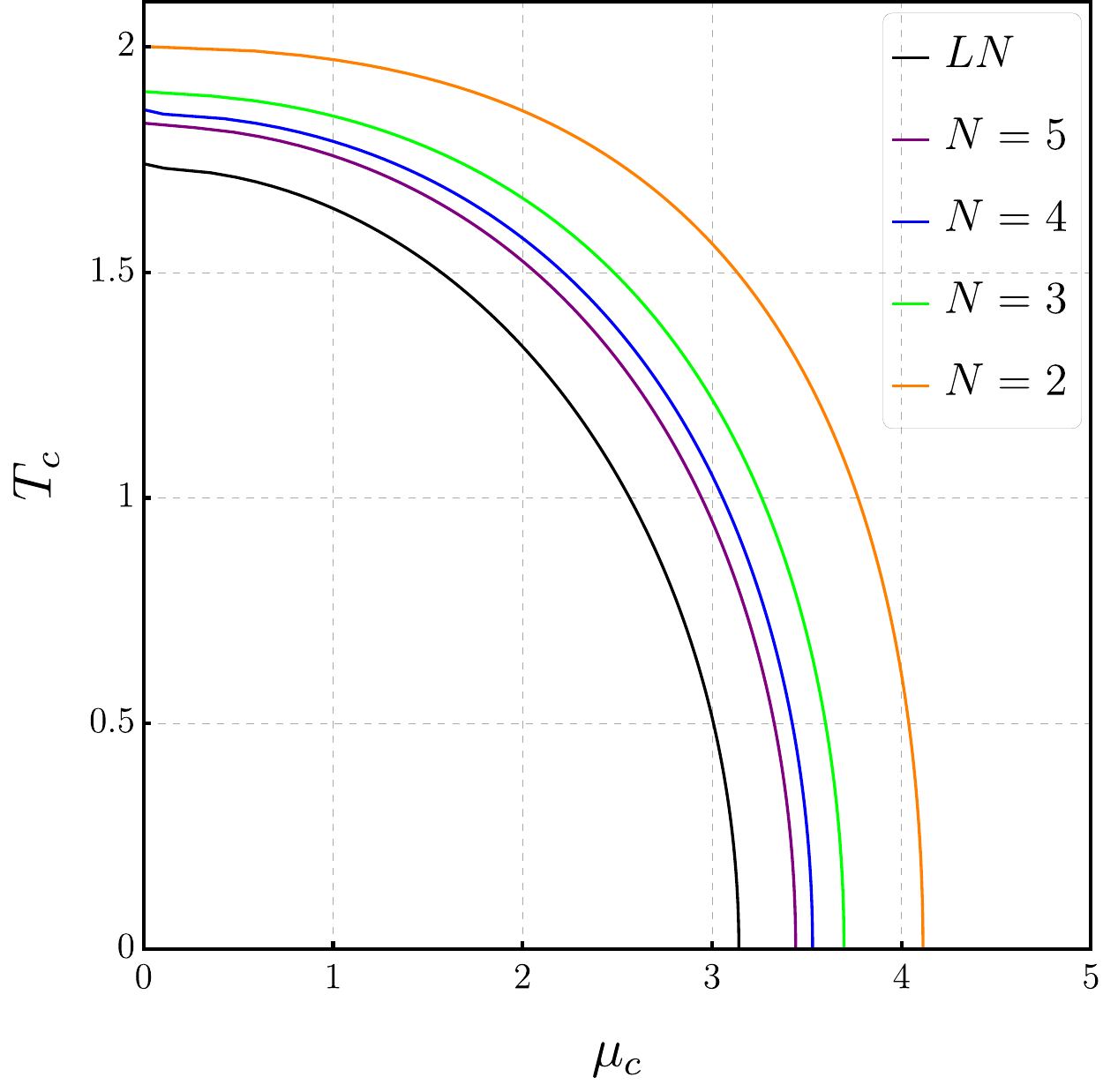}
\caption{The phase portrait of the critical temperature versus the critical
chemical potential in units of $g^{-1/2}~$. The blue and black lines correspond to the second-order phase transition. The black line sets the transition in the limit $N \rightarrow \infty$. }
\label{fig5}
\end{figure}
\begin{figure}[H]
\includegraphics[scale=0.3]{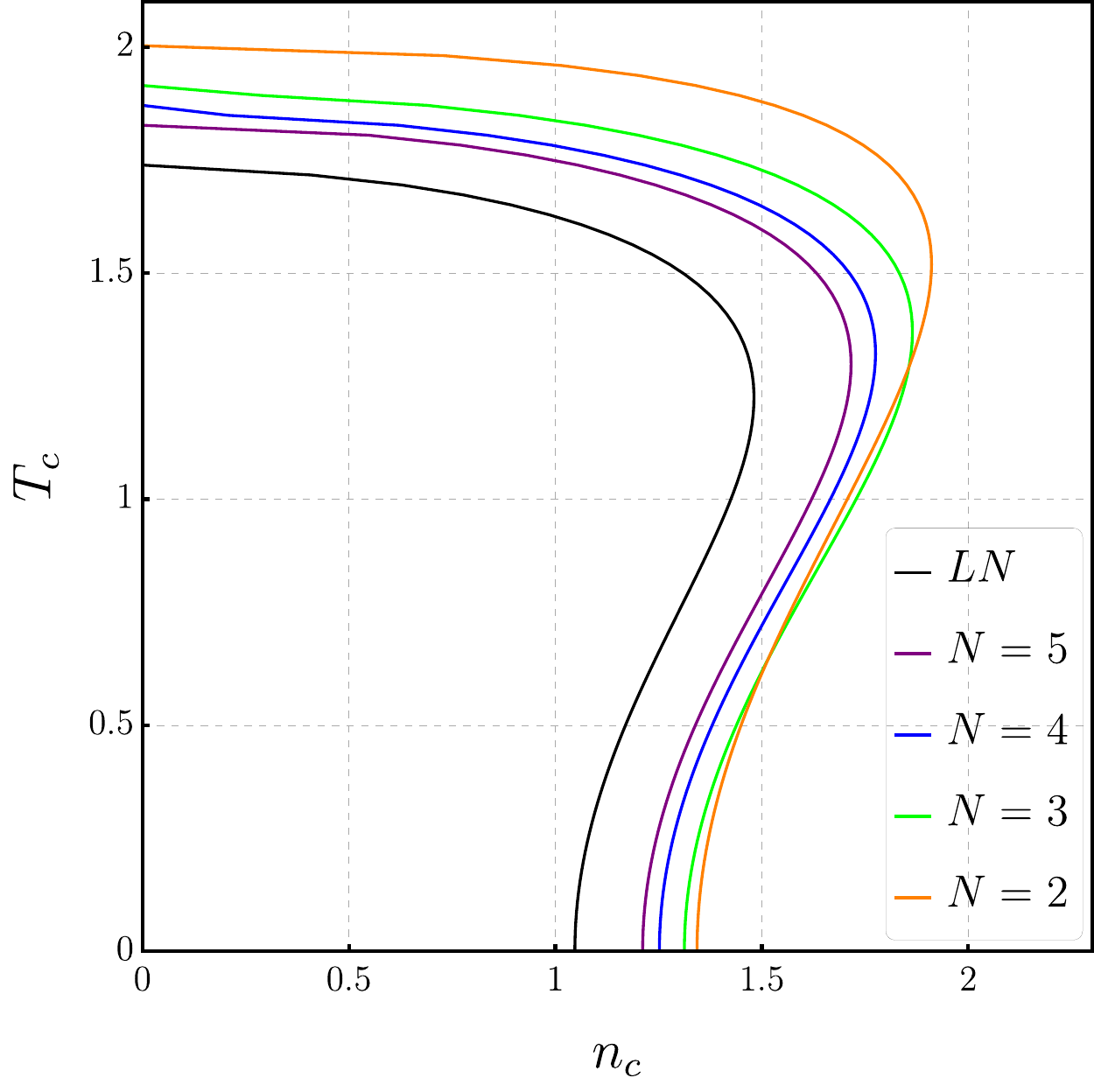}
\caption{Plot of the phase diagram of density $n_c \, \times \, T_c$, for $N= \{ 2,3,4,5 \}$, in units of $g^{-3/2}~$ and $g^{-1/2}~$, respectively. A black dashed line represents the second-order transition from the large-$N$ assumption.}
\label{fig6}
\end{figure}
\begin{figure}[H]
    \centering
    \includegraphics[width=.95\linewidth]{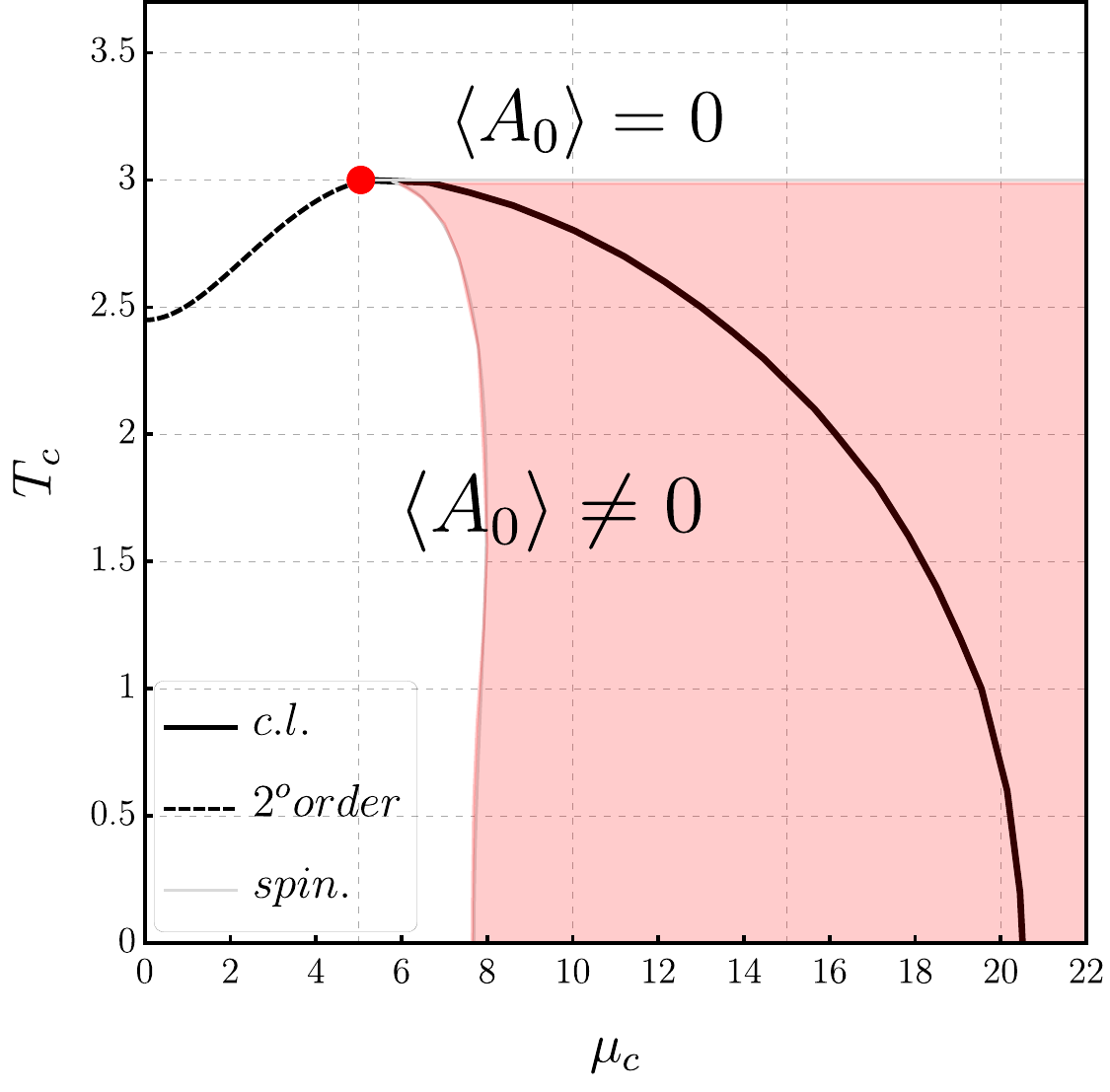}
    \caption{Plot of the phase diagram $\mu_c \, \times \, T_c$, for $N=1$, in units of $g^{-1/2}~$. The black dashed line represents the edge of the second-order phase transition. The solid line represents the coexisting line where the first-order phase transitions occur. The red dot represents the tricritical point. The light red region illustrates the metastable region, and the light gray line represents the spinodal curves. }
    \label{fig7}
\end{figure}
\begin{figure}[H]
    \centering   \includegraphics[width=.95\linewidth]{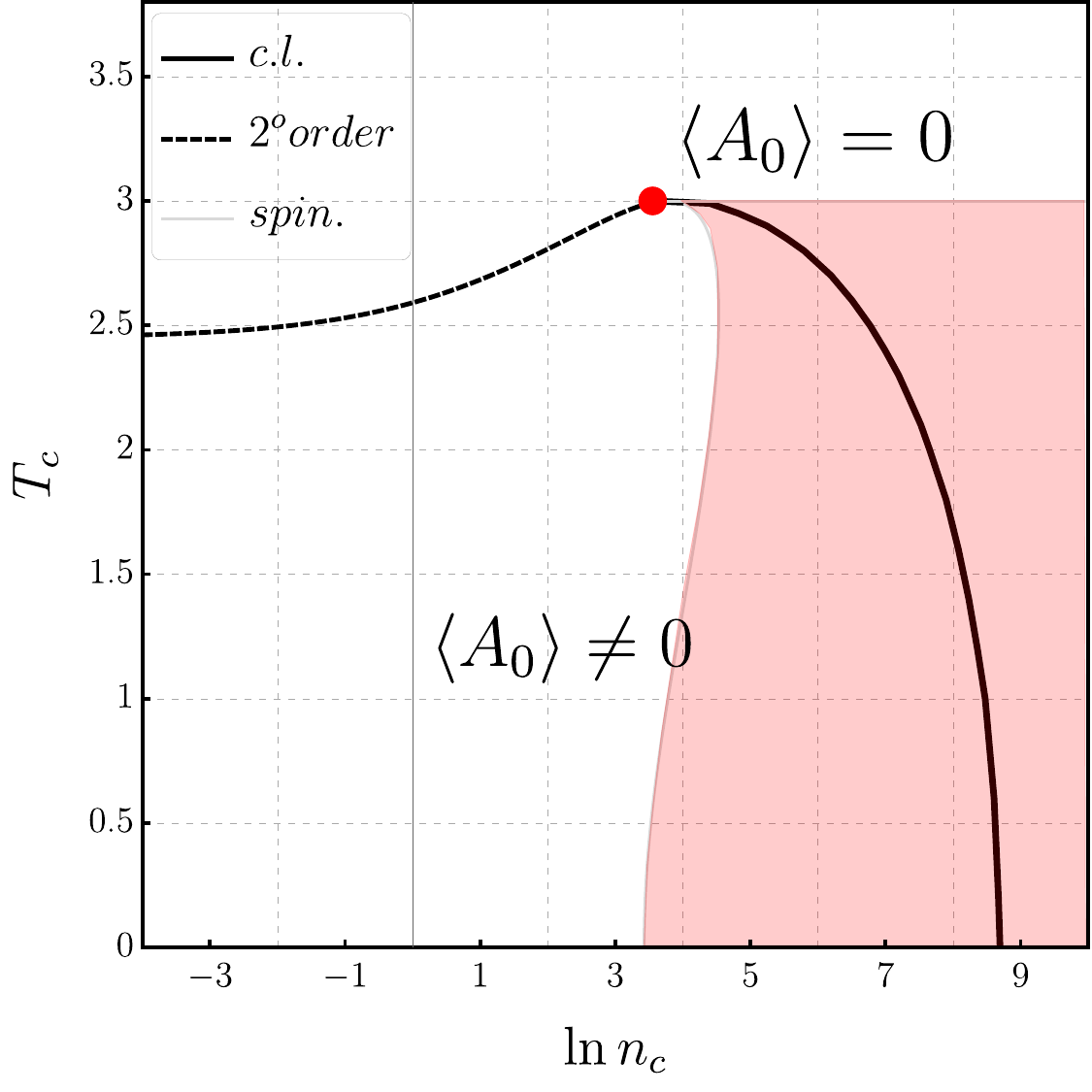}
    \caption{Plot of the phase diagram $ \log n_c \, \times \, T_c$, for $N=1$, in units of $g^{-3/2}~$ and $g^{-1/2}~$, respectively. The black dashed line represents the edge where the second-order phase transition occurs. The solid line represents the coexisting line where the first-order phase transitions occur. The red dot represents the tricritical point. The light red region represents the metastable region, and the light gray line represents the spinodal curves.  }
    \label{fig8}
\end{figure}
\section{SOME DISCUSSIONS OF THE RESULTS AND APPLICATIONS}\label{discuss}
Regarding Majorana spinors, it is widely recognized that the range of possible fermionic bilinears is limited due to Majorana conditions. For example, when dealing with one Majorana family, there are only three options: the scalar, pseudoscalar, and pseudo-vectorial (axial) bilinears. These bilinears can be used to model non-renormalizable quartic interactions like the axial version presented in this work. Utilizing both perturbative and non-perturbative methods, it can be demonstrated that these models provide a rich setting for dynamical symmetry breaking (DSB). Particularly, the case of axial self-interaction has been shown to reconcile the LSND and super-Kamiokande neutrino oscillation data \cite{gomes2022}. This phenomenon can be a first hint at new physics and has a strong connection with the results in this article. High-precision measurements of the neutrino oscillations can be a way to find stronger bounds on $G$ and a possible way to study flavor physics through the $N$ dependence of the axial condensate.

Going further, in the context of  Einstein-Cartan-Kibble-Sciama (ECKS) theory of gravity \cite{Poplawski:2010kb} which naturally extends general relativity to includes the spin of matter, the torsion of spacetime generates gravitational repulsion in the early Universe, preventing the cosmological singularity through a repulsive axial self-interactions between fermions. More general spinor dynamics in a theory where gravity is ruled by the Einstein-Cartan-Holst action can also generate attractive self-interactions \cite{Magueijo:2012ug}.  Interestingly, it implies that dark matter Majorana candidates, as shown in Ref. \cite{Neves:2021rlb}, could present first-order phase transitions at some densities. Furthermore, the astrophysical implications of these LSV bubbles formed by first-order phase transitions could impact the local dark matter distribution and be a new way to investigate the dark sector of the matter. 

About the chiral anomaly:
To keep the chiral anomaly controlled, an improvement of Eq. \eqref{lagr2} can be proposed as follows:
\begin{equation}\label{lagr3}
\mathcal{L} =
\overline{\psi}_k\left( i \slashed{\partial}- \slashed{A}\gamma_5 \right) \psi_k + \frac{N}{2 g} A_\mu A^\mu  + \lambda \left( \frac{g}{N}\partial_\mu A^\mu - \mathcal{A} \right) \; ,
\end{equation}
where $\lambda$ is a Lagrange multiplier and $\mathcal{A}= \partial \cdot j_5$ is the chiral anomaly, and 
$j^5_\mu=\bar{\psi}\gamma_\mu \gamma_5 \psi$ is the axial current. Integrating out the axial field $A_\mu$, one finds:
\begin{equation}\label{axion}
    \mathcal{L} =
\overline{\psi}_k( i \slashed{\partial}) \psi_k - \frac{g}{2 N} (j^5_\mu)^2  + \frac{1}{2}(\partial_\mu\bar{\lambda})^2+ \frac{\bar{\lambda}}{f_\lambda}( \partial \cdot j_5) \; ,
\end{equation}
with $\bar{\lambda} = \sqrt{\frac{3g}{N}}\lambda$ and $f_\lambda = \sqrt{\frac{3N}{g}}$.
It can be shown that Eq. \eqref{axion} is equivalent to Eq. \eqref{lagr1} only if $\bar{\lambda}= \text{constant}$. Thus, the $\bar{\lambda}$ field acts as an axionic field. Although this degree of freedom does not change the vacuum configurations found in this work, a detailed analysis of the excitations of the fields must be taken considering the axionic configurations.

Lastly, Weyl materials \cite{Armitage:2017cjs} is a fascinating area of study that draws connections between high-energy physics and condensed matter systems. Weyl fermions, initially theorized as massless Dirac fermions in relativistic quantum field theory, can emerge in solid materials when there is a break in either time-reversal or spatial-inversion symmetry. In these crystalline solids, the electronic energy bands can exhibit an emergent Lorentz symmetry at low energies, governing the dynamics of quasiparticle excitations above the ground state. When this symmetry is explicitly or spontaneously broken, it leads to the manifestation of various electronic phases within the system \cite{Kostelecky:2021bsb}. The results shown in this work can be adapted to study phase transitions on such materials.

\section{Conclusions}
\label{sec5}
In this paper, we propose the study of $N$ massless fermions based on the ref. \cite{gomes2022} in the presence
of temperature and chemical potential. The model has a fermionic quartic self-interaction that introduces an auxiliary
background $4$-vector $A^{\mu}$, and the finite-N corrections are implemented via the optimized perturbation theory (OPT)
via a regulator pseudo-vector field  $a^{\mu}$, the model dynamically breaks the Lorentz symmetry, generating an effective chiral chemical potential $\mu_5 = \langle A_0 \rangle$. Using the path integral approach of quantum field theory,
we obtain the effective potential of the model, in which the temperature is introduced via Matsubara formalism directly in the momentum integrals.
Thereby, we obtain the principle of minimal sensitivity (PMS) and the GAP equations in terms of the $4$-vector fields $A^{\mu}$ and $a^{\mu}$.

Looking ahead to the results, the vacuum solution for the gap equation given by $\bar{A}_0= \frac{\sqrt{3} \pi}{\sqrt{|g|}}\frac{1}{(1 - 1/2N)^{3/2}} $ is representative in terms to show the non-perturbative properties of the OPT approach, since the non-polynomial character of the
correction in comparison with the large-$N$ results. The non-perturbative results also show up on the critical temperature at zero chemical potential
$T_c =\frac{\sqrt3}{\sqrt{|g|}}\left( 1-1/2N\right)^{-1/2} $
 for $N \geq 1$ and the critical chemical potential at zero temperature given by
$\mu_c = \frac{\pi}{\sqrt{|g|}}\left( 1-5/6N \right)^{-1/2}$ for $N >1$. These non-perturbative results are one of our main results.

{Going further, the phase portraits are shown in Figs. \ref{fig5} ,\ref{fig6}, \ref{fig7} and \ref{fig6}, and one can see that, for $N>1$, the finite-$N$ corrections generate a region on the phase space $T \times \text{density}$ which is larger than the large-$N$ counterpart. Furthermore, one does not find any first-order phase transition, and the finite-$N$ correction maintains the second-order phase transition between the symmetric and LSV phases.}
  {Furthermore, one finds a very different system behavior in the case of $N=1$. In this case, the system transits in much richer forms between the symmetric and asymmetric phases. For small densities, one finds a second-order phase transition (same as for $N>1$). However, there is a small region within $T = (\sqrt6 g^{-1/2}, 3g^{-1/2})$ and $n\approx 0$, where the chiral balance is maintained, but the symmetry breaks when one increases the density. There is also a special region in the phase portrait represented by the light red region from Figs. \ref{fig7} and \ref{fig8}. One finds a metastable region where both the symmetric and LSV phases could coexist. In this region on the phase diagram, small fluctuations can generate LSV nucleations. The black solid curve from Figs. \ref{fig7} and \ref{fig8} represent the coexisting curve, where the phases coexists. We also obtain the tricritical point represented by the red dot in Figs. \ref{fig7} and \ref{fig8}, which gives us $\mu_{tric}\approx 5.05 g^{-1/2}$ ($n_{tric} \approx 35.2g^{-3/2}$) and $T_{tric}\approx 3 g^{-1/2}$. It is important to highlight that our result for $N=1$ shows that below the tricritical temperature and above the tricritical density, the OPT predicts a first-order phase transition region.}
In this work, the mass of the particles was neglected to extract analytical results. The Lagrangian \eqref{lagr1} may exhibit interesting
effects if a mass term is present. In particular, this could lead to a first-order phase transition   {for $N\geq 2$}, which warrants further investigation. Effects from $\delta^2$ corrections should also bring new information about the phase transition. { In ref. \cite{kneur2}, Kneur et al. have applied the OPT method on 2+1 systems up to $\delta^2$-order, showing the persistence of the first order transition and obtaining a small correction of $1.5\%$, compared to the $5\%$ correction of the $\delta^1$ counterpart. It shows that the OPT results at $\delta^2$ is a smaller correction than the order-$\delta^1$ results. Additionally, the contribution from $\delta^2$ generates terms that are momentum-dependent, such way more general non-perturbative techniques became necessary to explore the results. Since the first-order phase transition occurs for $N=1$, the robustness of this feature in $\delta^2$-expansion can be tested.} It is important to comment that if the effective Lagrangian has been calculated taking into account perturbations of $A^\mu= \bar{A}^\mu + \delta A^\mu(x)$ some sort of Carroll-Field-Jackiw term may appear. However, this structure does not appear only in constant vacuum configurations. These new features will be a target of forthcoming work.
\appendix
\section{Methodology of the regularized integrals}
\label{appA}
This appendix outlines the primary methodology used to derive the vacuum results, including the key steps involved in the process. By the use of the dimensional regularization with 't Hooft-Veltman prescription, one introduces the anticommutation relation (in $D$-dimensional space) given by $\{ \bar{\gamma}_\mu ,\bar{\gamma}_\nu\} = 2 \bar{\eta}_{\mu \nu}$ with $\bar{\eta}^{\mu \nu}\bar{\eta}_{\mu \nu}=D$. The D-dimensional Dirac matrices can be split as $\bar{\gamma}_{\mu} ={\gamma}_{\mu} + \hat{\gamma}_{\mu}$ ,  the four and $D-4$ dimensional subspaces, respectively.  The same separation can be applied in the metric $\bar{\eta}_{\mu \nu} = {\eta}_{\mu \nu}+\hat{\eta}_{\mu \nu}$ ,with ${\eta}^{\mu \nu}{\eta}_{\mu \nu}=4$,$\hat{\eta}^{\mu \nu}\hat{\eta}_{\mu \nu}=D-4$ and ${\eta}^{\mu \nu}\hat{\eta}_{\mu \nu}=0$.  The Dirac matrices respect the following algebra:
\begin{equation}
\{ {\gamma}_\mu ,{\gamma}_\nu\} = 2 {\eta}_{\mu \nu}~~,~~\{ \hat{\gamma}_\mu ,\hat{\gamma}_\nu\} = 2 \hat{\eta}_{\mu \nu}~~,~~ \{ {\gamma}_\mu ,\hat{\gamma}_\nu\} = 0 \; .
\end{equation}
Going further, the algebra of the Dirac matrices concerning the $\gamma_5$ is modified to:
\begin{equation}
    \{ {\gamma}_5 ,{\gamma}_\nu\} = 0~~,~~ [ \hat{\gamma}_\mu ,{\gamma}_5 ] = 0 \; .
\end{equation}
Therefore, one can calculate the integral as follows:
\begin{eqnarray}\nonumber
\mathcal{I}(\mathtt{a},D) &=& i\,\Lambda^{4-D} \!\! \int \frac{d^D\bar{p}}{(2 \pi)^D} \, \frac{1}{\slashed{\bar{p}}- \slashed{\mathtt{a}}\gamma_5}
\\
&&
\hspace{-1.5cm}
=i\,\Lambda^{4-D} \!\! \int \frac{d^D\bar{p}}{(2 \pi)^D} \, \frac{N_0(\bar{p})}{R(\bar{p})}  \; , \;\;
\end{eqnarray}
where $\bar{p}_\mu = p_\mu + \hat{p}_\mu$ and
$N_0(\bar{p}) = (\bar{p}^2+  \mathtt{a}^2 +2  \bar{p} \cdot \mathtt{a} \gamma_5 + [\slashed{\hat{p}},\slashed{\mathtt{a}}]\gamma_5)( \slashed{\bar{p}} + \slashed{\mathtt{a}} \gamma_5)$,
and the denominator is $R(\bar{p}) = (\bar{p}+\mathtt{a})^2(\bar{p}-\mathtt{a})^2 -4 \hat{p}^2 \mathtt{a}^2$. Thus:
\begin{eqnarray}
\mathcal{I}(\mathtt{a},D) &=& \approx i\,\Lambda^{4-D} \!\! \int \frac{d^D\bar{p}}{(2 \pi)^D} \times\\\nonumber
&&\hspace{-.5cm} \, \frac{(\bar{p}^2+  \mathtt{a}^2 +2  \bar{p} \cdot \mathtt{a} \gamma_5 + [\slashed{\hat{p}},\slashed{\mathtt{a}}]\gamma_5)( \slashed{\bar{p}} + \slashed{\mathtt{a}} \gamma_5)}{(\bar{p}+\mathtt{a})^2(\bar{p}-\mathtt{a})^2- 4 \hat{p}^2 \mathtt{a}^2}  \;  \;\; \\\nonumber~~.
\end{eqnarray}
Now, expanding the propagator to the first order on $\hat{p}$ one reaches:
\begin{eqnarray}\nonumber
\mathcal{I}(\mathtt{a},D)&&\approx \, \frac{ \mathtt{a}^2 (\slashed{\mathtt{a}}\gamma_5)}{72 \pi ^2}\left[3 \log \left(-\frac{\Lambda^2}{4  \mathtt{a}^2}\right)+\frac{3}{\epsilon }+8\right] \left(\bar{\eta}^{\mu\nu}\hat{\eta}_{\mu\nu} \right)\\
&&= -\frac{ \mathtt{a}^2 (\slashed{\mathtt{a}}\gamma_5)}{12 \pi ^2} + O(\epsilon) \; ,
\end{eqnarray}
where one uses the property $\hat{p}\cdot\mathtt{a}=0$ and $\bar{\eta}^{\mu\nu}\hat{\eta}_{\mu\nu}=D-4 = 2\epsilon$. Thus, one reaches:
\begin{equation}
\Pi^\mu(\mathtt{a},D) = \lim_{D \rightarrow 4} \mbox{Tr} \left[ \mathcal{I}(\mathtt{a},D)\bar{\gamma}_\mu \gamma_5 \right] = -\frac{ \mathtt{a}^2 \, \mathtt{a}^{\mu}}{3 \pi ^2} \; ,
\end{equation}
and
\begin{eqnarray}
\Xi(\mathtt{a},D) &=&  \lim_{D \rightarrow 4} \mbox{Tr} \left[ \mathcal{I}(\mathtt{a},D)\bar{\gamma}_\mu \gamma_5\mathcal{I}(\mathtt{a},D)\bar{\gamma}^\mu \gamma_5 \right]
\nonumber \\
&&=\frac{(\mathtt{a}\cdot\mathtt{a})^3}{18 \pi^4} \; .
\end{eqnarray}
Finally, one can find the expression to $\Omega(\mathtt{a})$ by solving the differential equation $\frac{\partial \Omega}{\partial \mathtt{a}_{\mu}} = - \Pi^\mu(\mathtt{a})$, and we obtain :
\begin{equation}
\Omega(\mathtt{a}) = \frac{ (\mathtt{a}\cdot\mathtt{a})^2 }{12 \pi ^2} + \text{constant} \; .
\end{equation}

\section{The Matsubara formalism }
\label{appB}
At $D=4$, The fermion propagator can be rewritten as follows :
\begin{eqnarray}
S({p}) &=& \frac{1}{\slashed{p} - \slashed{\mathtt{a}} \, \gamma_5}
\nonumber \\
&=& \frac{(p^2+\mathtt{a}^2 +2 p \cdot \mathtt{a}\,\gamma_5)(\slashed{p} + \slashed{\mathtt{a}}\gamma_5)}{(p^2-\mathtt{a}^2)^2+4[\,p^2 \, \mathtt{a}^2- (p \cdot \mathtt{a})^2 \,]}
\; , \hspace{0.8cm}
\end{eqnarray}
and through reorganization of the terms, the denominator can be rewritten as $D(p) = (p+\mathtt{a})^2(p-\mathtt{a})^2$.
For the study of the thermodynamic properties, one uses the Matsubara formalism, in which the $p^{0}$-the sum replaces momentum integration:
\begin{equation}
 \int \frac{d^{4}{p}}{(2\pi)^4} \rightarrow  i \, T \sum_{n=-\infty}^{\infty}  \int \frac{d^{3}{p}}{(2\pi)^{3}} \; ,
\end{equation}
where $p_0 = i\omega_n - \mu =i\,(2 n - 1) \, \pi \, T-\mu$. Based on the fact that the thermal contributions are finite, one can rewrite $\mathcal{I}(\mathtt{a})$ as follows:
\begin{eqnarray}\label{appres}\nonumber
\mathcal{I}(\mathtt{a}) &=& \!\! \int \frac{d^4p}{(2 \pi)^4} \, \frac{i}{\slashed{{p}}- \slashed{\mathtt{a}}\gamma_5}
\\
&&
\hspace{-1.5cm}
=\, \gamma_\mu \, \mathcal{I}_1^{\;\mu}(\mathtt{a}) + \gamma_\mu \, \gamma_5 \, \mathcal{I}_2^{\;\mu}(\mathtt{a}) \; ,
\end{eqnarray}
with
\begin{equation}
\mathcal{I}_1^{\; \mu}(\mathtt{a}) = \frac{i}{2} \sum_{\lambda=-1}^{+1} \int \frac{d^4p}{(2 \pi)^4} \frac{(p +\lambda \mathtt{a})^\mu}{(p+ \lambda \mathtt{a})^2} \; ,
\end{equation}
\begin{equation}
\mathcal{I}_2^{\; \mu}(\mathtt{a}) = \frac{i}{2} \sum_{\lambda=-1}^{+1} \lambda \int \frac{d^4p}{(2 \pi)^4}\frac{(p +\lambda \mathtt{a})^\mu}{(p+ \lambda \mathtt{a})^2} \; ,
\end{equation}
Going further, by use of the Matsubara mapping and using the following identity:

\begin{eqnarray}
 && T \sum_{n=-\infty}^{\infty} \frac{\omega_n + i \mu}{\left(\omega_n + i \mu \right)^2 + \omega_p^2}=
\nonumber \\
&&
 \frac{i}{2} \left[ \frac{1}{1+ e^ {(\omega_p + \mu)/T}} -  \frac{1}{1+ e^ {(\omega_p - \mu)/T} }	\right] \; ,
\end{eqnarray}

where $\omega_p=| {\bf p} |$ for a massless fermion. Thus, one can rewrite eq. \eqref{appres} as follows:

\begin{eqnarray}\nonumber
\mathcal{I}(\mathtt{a}) &=& \frac{1}{4} \sum_{\lambda=-1}^{+1} \, Y(\lambda \mathtt{a}_0) \, \gamma_0 \, \left(1 - \lambda \gamma_5 \right) \; ,
\end{eqnarray}
with

\begin{eqnarray}\label{f2}\nonumber
Y(\mu_\pm) \!&=&\!
\int \frac{d^{3}p}{(2 \pi)^{3}} \left[ \frac{1}{1+e^{\beta(|{\bf p}| + \mu_\pm)}} - \frac{1}{1+e^{\beta(|{\bf p}| - \mu_\pm)}}\right]
\nonumber \\
&&
\hspace{-1cm}
=-\frac{4T^3}{\pi^2} \left[ \, \text{Li}_3\left(-e^{\beta\mu_\pm }\right)- \text{Li}_3\left(-e^{-\beta\mu_\pm }\right) \, \right] \; ,
\;\;\;
\end{eqnarray}

with $\mu_\pm = \mu \mp \mathtt{a}_0$, and $\text{Li}_{3}$ is a Poly-logarithm function of third degree. Now, one can obtain the main results of the paper. One reaches:

\begin{equation}
\Pi^\mu(\mathtt{a}) = \mbox{Tr} \left[ \mathcal{I}(\mathtt{a})\gamma^\mu \gamma_5 \right] = \frac{1}{4} \, \delta^{\mu 0}\left[ Y(\mu_-) - Y(\mu_+) \right] \; ,
\end{equation}
\begin{equation}
    \Xi(\mathtt{a}) = -\mbox{Tr} \left[ \mathcal{I}(\mathtt{a})\gamma_\mu \gamma_5\mathcal{I}(\mathtt{a})\gamma^\mu \gamma_5 \right] = Y(\mu_-)^2+ Y(\mu_+)^2 \; .
\end{equation}
Finally,  $\Omega(\mathtt{a})$ is obtained by solving the differential equation $\frac{\partial \Omega}{\partial \mathtt{a}_{\mu}} = - \Pi_\mu(\mathtt{a})$.

%
%
\begin{acknowledgements}

YMPG would like to thank ROR, MBP, and GPB for the insightful discussions. YMPG is supported by a postdoctoral grant
from Fundação Carlos Chagas Filho de Amparo \`a Pesquisa do Estado do Rio de Janeiro (FAPERJ), grant No. E26/201.937/2020.

\end{acknowledgements}


\end{document}